\documentclass[twocolumn,preprintnumbers,amsmath,amssymb]{revtex4}

\usepackage{epsfig}
\usepackage{subfigure}
\usepackage{graphics}
\usepackage{amsmath}

\begin{document}

\title{Electrical Conductivity of Hot QCD Matter}

\author{W.~Cassing}
\affiliation{%
  Institut f{\"u}r Theoretische Physik, %
  Universit\"at Giessen, %
  35392 Giessen, %
  Germany %
}
\author{O.~Linnyk}%
\email{Olena.Linnyk@theo.physik.uni-giessen.de}
\affiliation{%
 Institut f{\"u}r Theoretische Physik, %
  Universit\"at Giessen, %
  35392 Giessen, %
  Germany %
}
\author{T.~Steinert}
\affiliation{%
  Institut f{\"u}r Theoretische Physik, %
  Universit\"at Giessen, %
  35392 Giessen, %
  Germany %
}
\author{V.~Ozvenchuk}
\affiliation{%
 Frankfurt Institute for Advanced Studies, %
 60438 Frankfurt am Main, %
 Germany %
}
\date{\today}
\begin{abstract}
We study the electric conductivity of hot QCD matter at various
temperatures $T$ within the off-shell Parton-Hadron-String Dynamics
(PHSD) transport approach for interacting partonic, hadronic or
mixed systems in a finite box with periodic boundary conditions. The
response of the strongly interacting system in equilibrium to an
external electric field defines the electric conductivity
$\sigma_0$. We find a sizable temperature dependence of the ratio
$\sigma_0/T$ well in line with calculations in a relaxation time
approach for $T_c \! < \! T < \! 2.5 \!\, T_c$. The ratio drops in
the hadronic phase with
$T$, shows a minimum close to $T_c$ and becomes approximately
constant ($\sim$0.3) above $\sim\!5T_c$. Our findings imply that the
QCD matter even at $T \! \approx \! T_c$ is a much better electric
conductor than $Cu$ or $Ag$ (at room temperature).
%
%
\end{abstract}
\maketitle
High energy heavy-ion reactions are studied experimentally and
theoretically to obtain information about the properties of nuclear
matter under the extreme conditions of high baryon density and/or
temperature. Ultra-relativistic heavy-ion collisions at the
Relativistic Heavy-Ion Collider (RHIC) and the Large Hadron Collider
(LHC) at CERN have produced a new state of matter, the strongly
interacting quark-gluon plasma (sQGP), for a couple of fm/c in
volumes up to a few thousand $fm^3$ in central reactions. The
produced QGP shows features of a strongly-interacting fluid unlike a
weakly-interacting parton gas \cite{StrCoupled1}
as had been expected from perturbative QCD (pQCD). Large values of
the observed azimuthal asymmetry of charged particles in momentum
space~\cite{STAR}
could quantitatively be well described by ideal hydrodynamics up to
transverse momenta of
1.5~GeV/c \cite{IdealHydro1}.
Recent studies of 'QCD matter' in equilibrium -- using lattice QCD
calculations \cite{l0,ll0} or partonic transport models in a finite
box with periodic boundary conditions \cite{Vitalii1,Vitalii2} --
have demonstrated that the ratio of the shear viscosity to entropy
density $\eta/s$ should have a minimum close to the critical
temperature $T_c$, similar to atomic and molecular systems
\cite{review,new1}. On the other hand, the ratio of the bulk
viscosity to the entropy density $\zeta/s$ should have a maximum
close to $T_c$ \cite{Vitalii2} or might even diverge at $T_c$
\cite{MaxBulk1}.
Indeed, the minimum of $\eta/s$ at $T_c \approx$ 160 MeV is close to
the lower bound of a perfect fluid with $\eta/s=
1/(4\pi)$~\cite{KSS} for infinitely coupled supersymmetric
Yang-Mills gauge theory (based on the AdS/CFT duality conjecture).
This suggests the `hot QCD matter' to be the `most perfect
fluid'~\cite{new1,new2,Barbara}. On the empirical side, relativistic
viscous hydrodynamic calculations (using the Israel-Stewart
framework) also require a very small $\eta/s$ of $0.08-0.24$ in
order to reproduce the RHIC elliptic flow $v_2$ data
\cite{ViscousHydro1};
these phenomenological findings thus are in accord with the
theoretical studies for $\eta/s$ in~\cite{Vitalii2,Mattiello,Greco}.

Whereas shear and bulk viscosities of hot QCD matter at finite
temperature $T$ presently are roughly known, the electric
conductivity $\sigma_0$ is another macroscopic quantity of
interest~\cite{Hirono:2012rt}. The basic question is: Is the 'hot
QCD matter' a good electric conductor? At first glance one might
expect the deconfined QCD medium to be highly conductive, since
color charges -- and associated electric charges of the fermions --
might move rather freely in the colored plasma. However, due to the
actual high interaction rates in the plasma -- reflected in a low
ratio $\eta/s$ -- this expectation is not so obvious. First results
from lattice calculation on the electromagnetic correlator provide
results that vary by more than an order of magnitude
\cite{l1,l2,l3,l4,l5}. Furthermore, the conductivity dependence on
the temperature $T$ (at $T\!\!>\!\!T_c$) is widely unknown, too.
The electric conductivity $\sigma_0$ is also important for the
creation of electromagnetic fields in ultra-relativistic
nucleus-nucleus collisions from partonic degrees-of-freedom, since
$\sigma_0$ specifies the imaginary part of the electromagnetic
(retarded) propagator and leads to an exponential decay of the
propagator in time $\sim \! \exp(-\sigma_0 (t-t')/({\hbar} c))$
\cite{Tuchin}.
High values of $\sigma_0$ would thus lead to the screening of
external electromagnetic fields in the bulk of the highly-conducting
quark-gluon plasma similar to the Meissner effect in
super-conductors as well as the ``skin-effect" for the electric
current.
Accordingly, a sufficient knowledge of $\sigma_0(T)$ is mandatory to
explore a possible generation of the Chiral-Magnetic-Effect (CME) in
predominantly peripheral heavy-ion reactions \cite{CME1}.

In this work we extract the electric conductivity $\sigma_0(T)$  for
`infinite parton/hadron matter' employing the Parton-Hadron-String
Dynamics (PHSD) transport approach \cite{PHSD1}, which is based on
generalized transport equations derived from the off-shell
Kadanoff-Baym equations \cite{Kadanoff1} for Green's functions in
phase-space representation (beyond the quasiparticle approximation). This approach
describes the full evolution of a relativistic heavy-ion collision
from the initial hard scatterings and string formation through the
dynamical deconfinement phase transition to the strongly-interacting
quark-gluon plasma (sQGP) as well as hadronization and the
subsequent interactions in the expanding hadronic phase. In the
hadronic sector PHSD is equivalent to the Hadron-String-Dynamics
(HSD) transport approach \cite{CBRep98} -- a covariant extension of
the Boltzmann-Uehling-Uhlenbeck (BUU) approach~\cite{Cass90} -- that
has been used for the description of $pA$ and $AA$ collisions from
lower SIS to RHIC energies in the past. On the other hand, the
partonic dynamics in PHSD is based on the Dynamical Quasi-Particle
Model (DQPM) \cite{DQPM1}, 
which describes QCD properties in terms of single-particle Green's
functions (in the sense of a two-particle irreducible (2 PI)
approach) and reproduces lattice QCD results -- including the
partonic equation of state -- in thermodynamic equilibrium. For
further details on the PHSD off-shell transport approach and
hadronization we refer the reader
to~\cite{PHSD1,Cassing,Bratkovskaya:2011wp,Vitalii1}.

Here, we concentrate on calculating the electric conductivity for
`infinite' QCD matter, which we simulate within a cubic
box with periodic boundary conditions at various values for the
energy density (or temperature). The size of the box is
fixed to $V\!=\!9^3$ fm$^3$ as in our previous investigations
\cite{Vitalii1,Vitalii2}. The initialization is done by populating
the box with light ($u,d$) and strange ($s$) quarks, antiquarks and
gluons slightly out of equilibrium. The system
approaches kinetic and chemical equilibrium during its evolution
within PHSD.  For more details on the simulation of equilibrated
partonic systems using PHSD in the box we refer the reader
to~\cite{Vitalii1}.

Let us remind that PHSD is an off-shell transport approach that
propagates quasi-particles with broad spectral functions.
Numerically, the continuous spectral distribution of mass of a
particle (given by its spectral function) is probed by a large
number of test-particles with (evolving) masses $M_j(t)$.
In order to include the effects from a constant external electric
field $E_z$, the propagation of each test-particle (in
$z-$direction) is performed with the additional force in the
equation of motion:
\begin{equation} \label{e1}
\frac{d}{dt} p_z^j = q_j e E_z,
\end{equation}
where $q_j$ denotes the fractional charges of the test-particles
($\pm 1/3, \pm 2/3$).
\begin{figure}
\begin{center}
\includegraphics[width=0.47\textwidth]{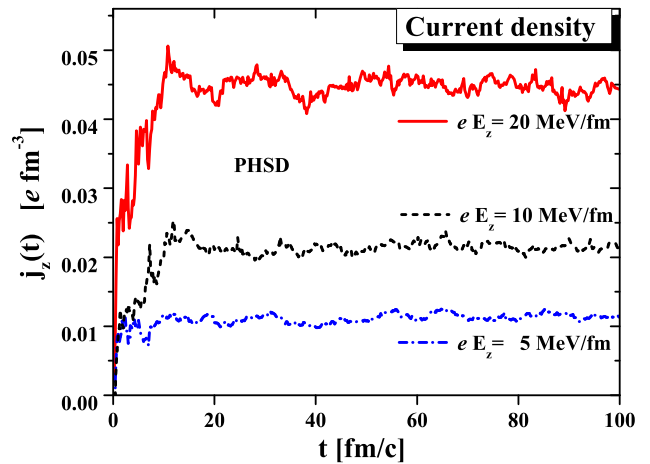}
\end{center}
\vspace{-0.6cm}
\caption{(color online) The electric current density $j_z(t)$
induced by an external (constant) electric field of strength $eE_z$
as a function of time $t$. The calculations are performed for a
system of partons at temperature $T= 190$~MeV in a box of volume
$729$~fm$^3$ within the PHSD approach. \vspace{-0.4cm}} \label{fig1}
\end{figure}
The  electric current density $j_z(t)$ then is given by
\begin{equation} \label{e2} j_z(t) = \frac{1}{V} \sum_j \ e q_j
\frac{p_z^j(t)}{M_j(t)}, \end{equation} where $M_j(t)$ is the mass
of the test-particle $j$ at time $t$. A note of caution has to be
given, since due to an external field we deal with an open system with increasing
energy density (temperature) in time. Therefore we employ sufficiently small
external fields, such that the energy increase during the computation
time stays below 2\% and the increase in temperature below 1
MeV.

\begin{figure}
\begin{center}
\includegraphics[width=0.47\textwidth]{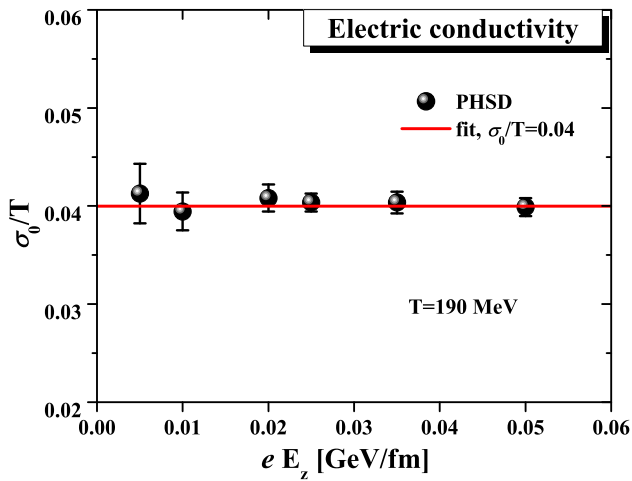}
\end{center}
\vspace{-0.7cm}
\caption{(color online) The ratio of the conductivity to the
temperature at $T$= 190 MeV as a function of the external electric
field $eE_z$. The statistical accuracy of the calculations is
reflected in the error bars. The straight line gives the best fit to
the dimensionless ratio $\sigma_0/T$. \vspace{-0.5cm} }\label{fig2}
\end{figure}

Fig.~\ref{fig1} presents the time dependence of the
electric current $j_z(t)$ induced by an external (constant)
electromagnetic field of strengths $eE_z = 5$~MeV/fm, $eE_z =
10$~MeV/fm and $eE_z = 20 $~MeV/fm (for the temperature  $T=
190$~MeV). It is seen that the current achieves an equilibrium value
(denoted by $j_{eq}$) that is proportional to the external field. In
fact, we obtain the conductivity from the ratio of the current
density and the electric field strength~\cite{footnote1}~\vspace{-10pt}
\begin{equation} \label{e4}
\frac{\sigma_0}{T} = \frac{j_{eq}}{E_z T},
\end{equation}
which is shown in Fig.~\ref{fig2} as a function of the external
field $e E_z$. All results are compatible with a constant ratio
(\ref{e4}) as indicated by the straight line. We note that our
numerical results for $\sigma_0$ do not depend on the volume $V$ of
the box within reasonable variations by a factor of~8.
%

\begin{figure}
\begin{center}
\includegraphics[width=0.48\textwidth]{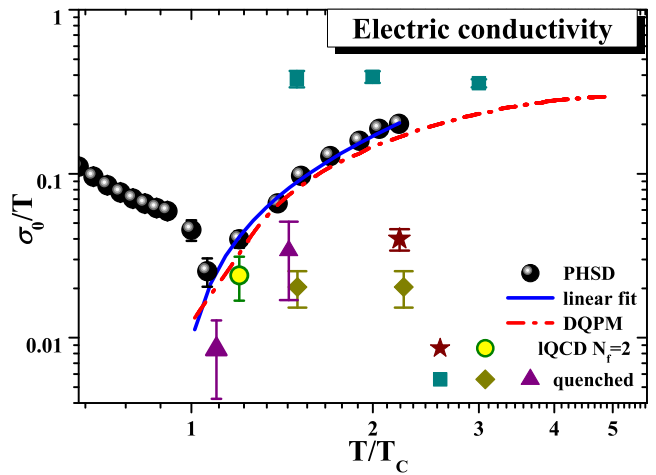}
\end{center}
\vspace{-0.7cm}
\caption{(color online) The ratio $\sigma_0/T$ as a function of the
scaled temperature $T/T_c$ ($T_c$ = 158 MeV). The full round symbols
show the PHSD results, the solid blue line is the linear fit to the
PHSD results (above $T_c$), while the dash-dotted red line gives the
corresponding ratio in the relaxation-time approach (employing the
DQPM parameters). The scattered symbols with error bars represent
the results from lattice QCD calculations:  triangles --
Refs.~\cite{l1}, diamonds -- Ref.~\cite{l2},  squares --
Ref.~\cite{l3},  star -- Ref.~\cite{l4}, open circle --
Ref.~\cite{l5}. We used for the average charge squared
$C_{EM}=8\pi\alpha/3$ with $\alpha=1/137$. Note that the pQCD result
at leading order beyond leading log~\cite{pQCD} is
$\sigma_0/T\approx5.97/e^2 \approx 65$.
 \vspace{-0.5cm}} \label{fig3}
\end{figure}

We have performed the PHSD studies for strongly interacting  systems
at various temperatures from $T=100$~MeV up to  $T=350$~MeV. The
respective results for the ratio $\sigma_0/T$ versus the scaled
temperature $T/T_c$ are displayed in Fig.~\ref{fig3} by the full
round symbols. We observe a decreasing ratio $\sigma_0/T$ with
$T/T_c$ in the hadronic phase, a minimum close to $T_c$ and an
approximately linear rise  with $T/T_c$ above $T_c$ (=158 MeV).
Within the error bars of our calculations (which in~Fig.~\ref{fig3}
are indicated by the size of the symbols above $T_c$), the
conductivity in the partonic phase is described by
\begin{equation} \label{e6} \frac{\sigma_0(T)}{T} \approx 0.01 +
0.16 \frac{T-T_c}{T_c}
\end{equation}
for $T_c\! \le \! T \! \le \! 2.2T_c$. The lQCD
numbers~\cite{l1,l2,l3,l4,l5} are represented by symbols with error
bars (using $C_{EM}=2 e^2/3$, $e^2=4\pi\alpha$, $\alpha=1/137$).
In view of the pQCD prediction of a constant asymptotic value for
$\sigma_0/T \approx 5.9769/e^2 \approx65$ in leading order of the
coupling~\cite{l1,Olena13,pQCD},
%
%
%
such a linear rise of the ratio with temperature might be
surprising, but it can be understood in simple terms as demonstrated
below.
Only at the highest temperatures studied ($\sim 5 T_c$) we will see
a stabilization of the ratio.

We recall that the electric conductivity of gases, liquids and solid
states is described in the relaxation time approach by~\vspace{-10pt}
\begin{equation} \label{eq7} \sigma_0 = \frac{e^2 n_e \tau}{m_e^*} ,
\end{equation}
where $n$ denotes the density of non-localized charges, $\tau$ is
the relaxation time of the charge carriers in the medium and $m_e^*$
their effective mass. This expression can be directly computed for
partonic degrees-of-freedom within the DQPM, which was used to match
in PHSD the quasiparticles properties to lattice QCD results in
equilibrium for the equation-of-state (EoS) as well as various
correlators \cite{DQPM1}.
We note that the electromagnetic correlator from lQCD calculations
\cite{l1} appears to match rather well the back-to-back dilepton
rate from PHSD at $T=1.45 T_c$ (cf. Fig.~2 in~\cite{Olena13}), which
suggests that the results of our calculations for $\sigma_0$ should
also be close to the lQCD extrapolations from~\cite{l1}.

%
%

In the DQPM, the relaxation time for quarks/antiquarks is given by
$\tau = 1/\Gamma_q(T)$~\cite{footnote1}, where $\Gamma_q(T)$ is the
width of the quasiparticle spectral function
(cf.~\cite{DQPM1,Bratkovskaya:2011wp}). Furthermore, the spectral
distribution for the mass of the quasiparticle has a finite pole
mass $M_q(T)$ that is also fixed in the DQPM, as well as the density
of ($u, \bar{u}, d, \bar{d}, s, \bar{s}$) quarks/antiquarks as a
function of temperature (cf.~\cite{DQPM1,Bratkovskaya:2011wp}).
Thus, we obtain for the dimensionless ratio (\ref{e4}) the
expression~\cite{footnote1}
\begin{equation} \label{e8}
\frac{\sigma_0 (T)}{T} \approx \frac{2}{9} \frac{e^2 n_q(T)}{M_q(T)
\Gamma_q(T) T} ,
\end{equation}
where $n_q(T)$ denotes the total density of quarks and antiquarks
and the prefactor $2/9$ reflects the flavor averaged fractional
quark charge squared $(\sum_f q_f^2)/3$. The result for the ratio
(\ref{e8}) is displayed in Fig.~\ref{fig3} (dash-dot line) and does
not involve any new parameters. Apparently, the PHSD results in
equilibrium and the relaxation-time estimates match well up to
$\sim$2$T_c$, which demonstrates again that PHSD in equilibrium is
a proper transport realization of the DQPM~\cite{Vitalii1}.

Our results from the DQPM suggest that above $T\sim 5 T_c$ the
dimensionless ratio (\ref{e8}) becomes approximately constant
($\approx0.3$). This comes about as follows: At high temperature $T$
the parton density scales as $\sim T^3$, while $M_q(T) \sim T$ and
$\Gamma_q(T) \sim T$. Accordingly the ratio (\ref{e8}) is
approximately constant.
Note, however, that energy densities
corresponding to $T
> 5 T_c$ are not reached in present experiments with
heavy-ions at RHIC or LHC!
On the other hand, $\sigma_0/T$ rises with decreasing temperature
below $T_c$ (in the dominantly hadronic phase), because at lower
temperatures the system merges to a moderately interacting system of
pions, which in view of Eq. (5) has a larger charge (squared) to
mass ratio than in the partonic phase as well as a longer relaxation
time.

In summary, we have evaluated the electric conductivity
$\sigma_0(T)$ of the quark-gluon plasma as well as the hadronic
phase  as a function of temperature $T$ by employing the
Parton-Hadron-String Dynamics (PHSD) off-shell transport model in a
finite box for the simulation of dynamical partonic, hadronic or
mixed systems in equilibrium. The PHSD approach in the partonic
sector is based on the lattice QCD equation of state of~\cite{lQCD};
accordingly, it describes the QGP entropy density $s(T)$, the energy
density $\varepsilon(T)$ and the pressure $p(T)$ from
lQCD~\cite{PHSD1,Bratkovskaya:2011wp,Vitalii1} very well. Studies of
the QCD matter within PHSD have previously given reasonable results
also for the shear and bulk viscosities $\eta$ and $\zeta$ versus
$T$~\cite{Vitalii2}. We find in the present study that the
dimensionless ratio $\sigma_0/T$ rises above $T_c$ approximately
linearly with $T$ up to $T=2.5 T_c$, but approaches a constant above
$5 T_c$, as expected from pQCD. This finding is naturally explained
within the relaxation-time approach using the DQPM spectral
functions. Below $T_c$ the ratio $\sigma_0/T$ rises with decreasing
temperature because the system merges to a moderately interacting
gas of pions with a larger charge to mass ratio than in the partonic
phase and a longer relaxation time.

The actual values for the electric conductivity $\sigma_0(T)$ show
that the sQGP even at its minimum (at $T \approx T_c$) is a better electric
conductor than $Cu$ or $Ag$  (at room temperature) by about a factor
of 500 (using $2\pi \hbar/e^2=2.58 \cdot 10^4
\Omega$~\cite{footnote2}). Furthermore, the damping of the
electromagnetic propagator in the partonic medium $\tau_e(T) =
(\hbar c)/\sigma_0(T)$ is moderate for temperatures $T \le 2 T_c$,
yet becomes shorter than 2~fm/c  for $T > 2.8 T_c$.
This suggests the potential importance of the 'skin-effect' in the
response of the partonic medium as created in the mid-rapidity
region of heavy-ion collisions at the LHC to the electromagnetic
fields generated by the spectators.

V.O. acknowledges the financial support from the HGS-Hire, H-QM and
the LOEWE center HICforFAIR.
\vspace{-0.5cm}


\begin{thebibliography}{99}
\bibitem{footnote1} Note that one has to use the conversion constant  $({\hbar}
c)$ in order to translate the volume, electric field strength, etc
to  natural units (GeV) in the calculation of the ratio.
\bibitem{footnote2} I.e. the explicit conversion
of the conductivity units between the SI and the natural units is
given by
%
1~$\Omega^{-1}$m$^{-1}=7.39\cdot10^{-5} \mbox{ eV } e^2 /\hbar
/(\hbar c)/(4\pi \alpha)$.
\end{thebibliography}
\end{document}